# Strong impact of grain boundaries on the thermoelectric properties of non-equilibrium synthesized p-type $Ce_{1.05}Fe_4Sb_{12.04}$ filled skutterudites with nanostructure

Qing Jie[1,2], Juan Zhou[1,2], Xun Shi[3], Ivo K. Dimitrov[1] and Qiang Li[1*]

[1]Condensed Matter Physics and Materials Science Department, Brookhaven National Laboratory, Upton, New York 11973, USA

[2]Department of Materials Science and Engineering, Stony Brook University, Stony Brook, New York 11794, USA

[3]CAS Key Laboratory of Materials for Energy Conversion, Shanghai Institute of Ceramics, Chinese Academy of Sciences, Shanghai, 200050 China

**ABSTRACT**

p-type $Ce_{1.05}Fe_4Sb_{12.04}$ filled skutterudites with much improved thermoelectric properties have been synthesized by rapidly converting nearly amorphous ribbons into crystalline pellets under pressure. It is found that this process greatly suppresses grain growth and second phase formation/segregation, and hence results in the samples consisting of nano-sized grains with strongly-coupled grain boundaries, as observed by transmission electron microscopy. The room temperature carrier mobility in these samples is significantly higher (nearly double) than those in the samples of the same starting composition made by the conventional solid-state reaction. Nanostructure reduces the lattice thermal conductivity, while cleaner grain boundaries permit higher electron conduction.

Widespread application of thermoelectricity relies on the availability of superior materials that would exceed the performance of current state-of-the-art thermoelectric (TE) materials, such as $Bi_2Te_3$-$Sb_2Te_3$, PbTe, and Si-Ge solid solutions. The efficiency of a TE material is related to the dimensionless figure of merit $ZT$, expressed as $ZT = S^2T/(\rho\kappa)$, where $S$, $\rho$, $\kappa$, $T$ are the Seebeck coefficient, electrical resistivity, thermal conductivity and absolute temperature, respectively. Filled skutterudites are among the most promising material systems for intermediate temperature power generation owing to both good thermoelectric and mechanical properties.[1-4] Extensive work has been performed on synthesizing skutterudite-based materials with the primary goal of lowering their thermal conductivity by introducing "rattling" atoms into the voids in the structure. In 1995, Morelli and Meisner reported a near tenfold reduction in the thermal conductivity by filling the voids in $Fe_4Sb_{12}$ based skutterudites with Ce atoms.[5] On the n-type side, Nolas et al. reported reductions in the thermal conductivity of Yb filled skutterudites, by as much as 50%,[6,7] and hence the $ZT$ value of $Yb_{0.2}Co_4Sb_{12}$ was raised above 1 at temperatures higher than 600K. Recently, the thermal conductivity of skutterudites has been further reduced by filling the voids with two or three different elements.[8,9] Despite the successful reduction of $\kappa$, the carrier mobility has also been degraded rather strongly as a consequence of filling in some cases,[5,6,10] leading to a decrease in the power factors of those materials.

Nanostructuring has been shown to be an alternative method for reducing the lattice thermal conductivity in bulk materials. Poudel *et al.* successfully reduced the thermal conductivity in p-type BiSbTe alloy by producing nano-sized grains via a special ball milling technique, and raised the peak ZT value from 1 to 1.4.[11] Recently, a non-equilibrium synthesis technique, which employs melt-spinning with subsequent spark plasma sintering (SPS), has been introduced in thermoelectric research. Filled skutterudites synthesized via this method exhibit significantly reduced thermal conductivity.[12-14] However, detailed studies on the relationship between electrical conduction properties and structure in the latter materials are limited. In this work, we prepared p-type $Ce_{1.05}Fe_4Sb_{12.04}$ filled skutterudites by both non-equilibrium (melt-spinning followed by SPS) and equilibrium (long term annealing at high temperature) process. In order to understand the origin of the observed thermoelectric property improvement, we carried out a coordinated structural and transport investigation. We found that higher carrier mobility in non-equilibrium processed samples can be directly related to their much improved grain boundary quality. This is the result of the rapid conversion process that limits the grain growth and impurity phase segregation.

The filled skutterudite ingots were prepared by melting the mixture of stoichiometric amounts of high purity Ce (99.8% min), Fe (99.98%) and Sb (99.9999%) above 1050 ˚C before quenching the mixture in water. A part of a quenched ingot was used for melt spinning. The melt-spun ribbons were ground and sintered into pellets under 50 MPa pressure using a SPS machine at 620˚C for 2 minutes, and the resulting samples were labeled as MS. For comparison, another part of the quenched ingot was annealed at 700˚C for 30h to form the single phase compound, followed by powdering and SPS sintering into pellets (labeled as AN) using the same pressure, temperature and time as those used for the MS samples. X-ray powder diffractometry investigation shows both sintered pellets predominantly comprise of a single filled skutterudite phase (less than 5% impurity phases). Interestingly, the MS samples appear to have fewer impurities, suggesting that this direct rapid conversion process is very effective.

Fig. 1(a) shows the fracture surface of the MS sample after SPS, whereas the inset is the HRTEM image showing the structure of the melt-spun ribbon prior to SPS. Nano-sized crystals are found in the amorphous matrix of the melt-spun ribbons, as the result of rapid solidification. In the subsequent 2 minutes of SPS, the amorphous materials were completely converted into crystalline phase and limited grain growth is observed, that results in highly dense MS pellets (above 99% theoretical density) with an average grain size of ~ 300 nm. In contrast, the grain size of AN sample (tens of μm), as shown in Fig. 1 (b), is more than one order of magnitude bigger than that in the MS sample. Remarkably, the fracture surface of the MS sample goes along its grain boundaries, while the fracture surface of the AN sample prefers to go through its grains. Hence, the MS samples are likely to have better fracture toughness. In general, the longer the distance a

The microstructure of melt spun ribbons and sintered samples were examined using scanning electron microscope (SEM) and high resolution transmission electron microscope (HRTEM). Thermopower and thermal conductivity were measured in a Quantum Design Physical Property Measurement System with the Thermal Transport Option. The resistivity and Hall coefficient were measured using a standard 4-probe method. Lattice thermal conductivities ($\kappa_L$) were obtained by subtracting the electronic contribution ($\kappa_e$) from the total thermal conductivity $\kappa$. The electronic thermal conductivity was derived by the Wiedemann Franz law: $\kappa_e = LT\sigma$, where $L$ is the Lorenz number (A moderated $L = 2.0 \times 10^{-8}$ WΩK$^{-2}$ is used here).

crack propagates, such as going along a grain boundary rather than going through a grain, the more energy it consumes.

Fig. 1(c) and (d) show the typical results of HRTEM characterization of grain boundaries in both the MS samples and the AN samples. Well-coupled grains with structurally intact grain boundaries are mostly found in the MS samples, while poorly coupled grains with second phase at grain boundaries are mostly found in the AN samples. The width of this second phase region ranges from 1 to 10 nm. By using energy dispersive spectroscopy (EDS), we determined the atomic ratio of the component elements at the grain boundaries and within the grains of the AN samples. These grain boundaries are Ce-rich (as much as 5-10 times higher than the starting composition). For example: the Ce:Fe:Sb ratio in two of the grain boundaries is 58.5 : 1.5 : 40 and 38.9 : 8.1 : 53 respectively, while the compositions within each grain are very close to the stoichiometric ratio. It is apparent that long term annealing necessary for the preparation of AN samples are also causing impurity or filler atoms to segregate at grain boundaries. In contrast, the non-equilibrium process produces much cleaner grain boundaries in the MS samples.

Fig. 2 shows the temperature dependence of lattice thermal conductivity $\kappa_L$ for the MS and AN samples. As expected, the MS samples with nano-sized grains have much lower $\kappa_L$ than the AN samples, and a reduction by half is observed at $T\sim50K$. Fig. 3 compares the temperature dependences of $\rho$, $S$, power factor and $ZT$ for MS and AN samples. Both S and ρ curves of these two samples cross at 80K. Below this crossing temperature, the MS sample has higher $\rho$ and lower $S$ than the AN sample. Above this temperature, the resistivity of the MS samples becomes significantly lower than that of the AN samples. Above 150 K. this resistivity difference maintains the same level with the increase of temperature, while the thermopower of both samples are virtually the same (within 5%). With lower lattice thermal conductivity and higher power factor, the ZT of the MS sample is higher than that of the AN sample over the entire temperature region, as shown in Fig. 3d.

Fig. 4 displays the temperature dependences of the hole concentration ($p$, the inset) and hole mobility ($\mu$) in the MS and AN specimens. Both of them have relatively high carrier concentrations ($\sim 10^{21}/cm^3$), with a temperature dependence similar to previous observation in this system.[15] Remarkably, at room temperature, the hole mobility of the MS sample is about twice as high as that of the AN sample, while its carrier concentration stays lower. This indicates that the low resistivity observed in the MS samples at elevated temperatures (> 80K) can be mainly attributed to its high mobility.

We found that the temperature dependence of carrier mobility above 80 K follow the relation $\mu \propto T^{-\frac{1}{2}} \exp(-\frac{E_b}{k_B T})$ very well, where $E_b$ is the boundary potential barrier height,[16] and $k_B$ is the Boltzman constant. The solid lines shown in Fig. 4 are the corresponding fitting curves with fitting parameters $E_b \sim$ 187K (~ 16 meV) for AN samples and $E_b \sim$ 90K (~ 7.8 meV) for MS samples. The derived energy barriers are qualitatively consistent with the results of structural characterization. This is because the thicker and dirty grain boundary in AN samples is capable to trap more carriers and build up a higher potential barrier. The overall temperature dependence of the mobility behavior is explained in the followings. At low temperatures (<80K), the kinetic energy of electrons is much lower than the energy barrier height so that the number of barriers dominates the transport behavior. Higher density of grain boundaries (more scattering events) lead to lower mobility and hence higher resistivity in the nano-structured MS samples. As temperature increases, the phonon scattering becomes important and leads to the quick decrease of carrier mobility in both samples. On the other hand, the kinetic energies of electrons increase and get close to the barrier height ($E_b \sim$ 90K) at T >80 K in the MS sample, giving rise to the higher hole mobility and lower resistivity observed in the MS sample. The 80 K is more or less a crossover temperature.

In summary, we have shown that non-equilibrium synthesis is a fast and effective method for preparing high performance thermoelectric filled skutterudites. This rapid conversion process produces nano-sized grains with cleaner grain boundaries. In comparison, the conventional long term annealing method produces large grains with dirty grain boundaries, where the filler atoms and second phases segregate. Combined structural and transport measurements provide clear evidence that the grain boundaries play a major role in the electron scattering in these materials. The temperature dependence of the carrier conduction behavior is explained in terms of the crossover from the dominance of the energy barrier density at low temperature to the energy height at elevated temperature.

*Contacting author: qiangli@bnl.gov

**ACKNOWLEDGEMENTS**

This work was in part supported by the U.S. Department of Energy, Office of Basic Energy Science, under Contract No. DE-AC02-98CH10886. We thank the Center for Functional Nanomaterials, Brookhaven National Laboratory for generous support in using its facilities.

**REFERENCES**

1. B. C. Sales, D. Mandrus, and R. K. Williams,  Science **272** (5266), 1325 (1996).
2. B. C. Sales, D. Mandrus, B. C. Chakoumakos, V. Keppens, and J. R. Thompson, Phys. Rev. B **56** (23), 15081 (1997).
3. Alex Borshchevsky Jean-Pierre Fleurial, Thierry Caillat, donald T. Morelli and Gregory P. Meisner, Fifteenth International Conference on Thermoelectrics, 91 (1996).
4. J. R. Salvador, J. Yang, X. Shi, H. Wang, A. A. Wereszczak, H. Kong, and C. Uher,  Philos. Mag. **89** (19), 1517 (2009).
5. D. T. Morelli and G. P. Meisner,  J. Appl. Phys. **77** (8), 3777 (1995).
6. G. S. Nolas, J. L. Cohn, and G. A. Slack,  Phys. Rev. B **58** (1), 164 (1998).
7. G. S. Nolas, M. Kaeser, R. T. Littleton, and T. M. Tritt, Appl. Phys. Lett. **77** (12), 1855 (2000).
8. X. Shi, H. Kong, C. P. Li, C. Uher, J. Yang, J. R. Salvador, H. Wang, L. Chen, and W. Zhang,  Appl. Phys. Lett. **92**, 182101 (2008).
9. X. Shi, J. R. Salvador, J. Yang, and H. Wang,  J. Electron. Mater. **38** (7), 930 (2009).
10. D. T. Morelli, G. P. Meisner, B. X. Chen, S. Q. Hu, and C. Uher,  Phys. Rev. B **56** (12), 7376 (1997).

[11]  B. Poudel, Q. Hao, Y. Ma, Y. C. Lan, A. Minnich, B. Yu, X. Yan, D. Z. Wang, A. Muto, D. Vashaee, X. Y. Chen, J. M. Liu, M. S. Dresselhaus, G. Chen, and Z. Ren,  Science **320** (5876), 634 (2008).

[12]  H. Li, X. F. Tang, X. L. Su, and Q. J. Zhang,  Appl. Phys. Lett. **92**, 202114 (2008).

[13]  H. Li, X. F. Tang, Q. J. Zhang, and C. Uher,  Appl. Phys. Lett. **93**, 252109 (2008).

[14]  Q. Li, Z. W. Lin, and J. Zhou,  J. Electron. Mater. **38** (7), 1268 (2009).

[15]  B. X. Chen, J. H. Xu, C. Uher, D. T. Morelli, G. P. Meisner, J. P. Fleurial, T. Caillat, and A. Borshchevsky,  Phys. Rev. B **55** (3), 1476 (1997).

[16]  J. Y. W. Seto,  J. Appl. Phys. **46** (12), 5247 (1975).

**FIGURE LEGENDS**

Figure 1. SEM images of the fracture surface of (a) non-equilibrium MS and (b) equilibrium AN samples. The inset to (a) is the HRTEM image showing the nanostructure of the melt-spun ribbon prior to SPS. HRTEM images showing typical grain boundaries in the MS samples (c) and the AN samples (d).

Figure 2. Temperature dependence of lattice thermal conductivity $\kappa_L$ in samples prepared by non-equilibrium (red circle) and equilibrium (black square) method.

Figure 3. The temperature dependence of electrical resistivity (a), thermopower (b), power factor ($PF=S^2/\rho$) (c) and $ZT$ (d) in samples prepared by non-equilibrium (red circle) and equilibrium (black square) method.

Figure 4. The temperature dependence of hole mobility and hole concentration (the inset) in p-type $Ce_{1.05}Fe_4Sb_{12.04}$ filled skutterudites prepared by non-equilibrium (red circle) and equilibrium (black square) method. The hole mobilities of two samples were calculated using equation $\sigma = pe\mu$, where $e$ is the electric charge of the carrier. The lines are corresponding fitting curves (see text).

**FIGURES**

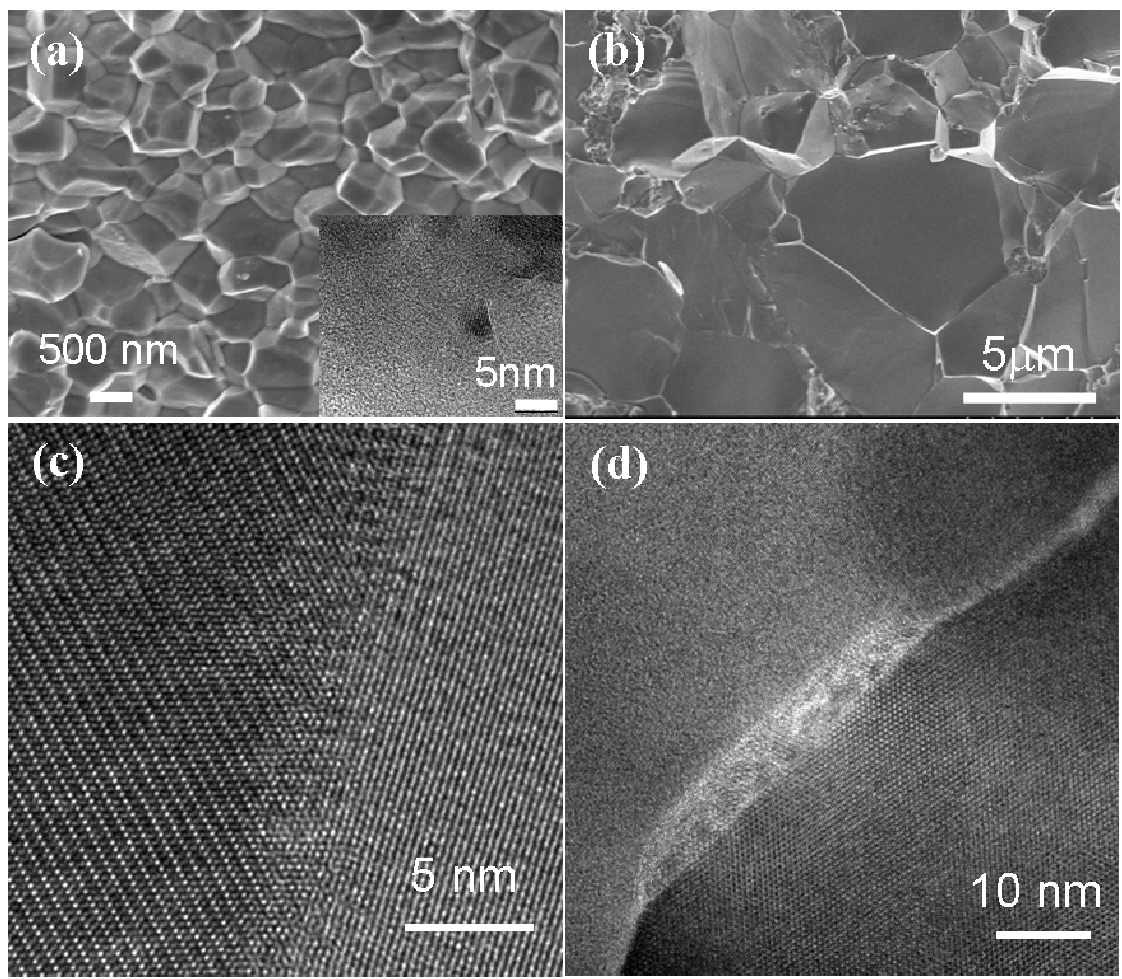

Figure 1

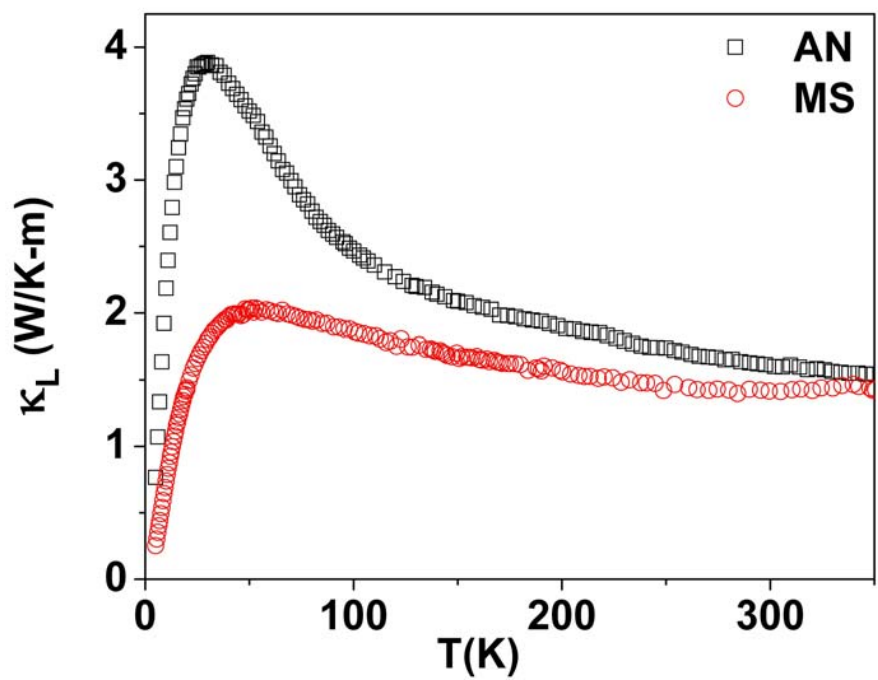

Figure 2

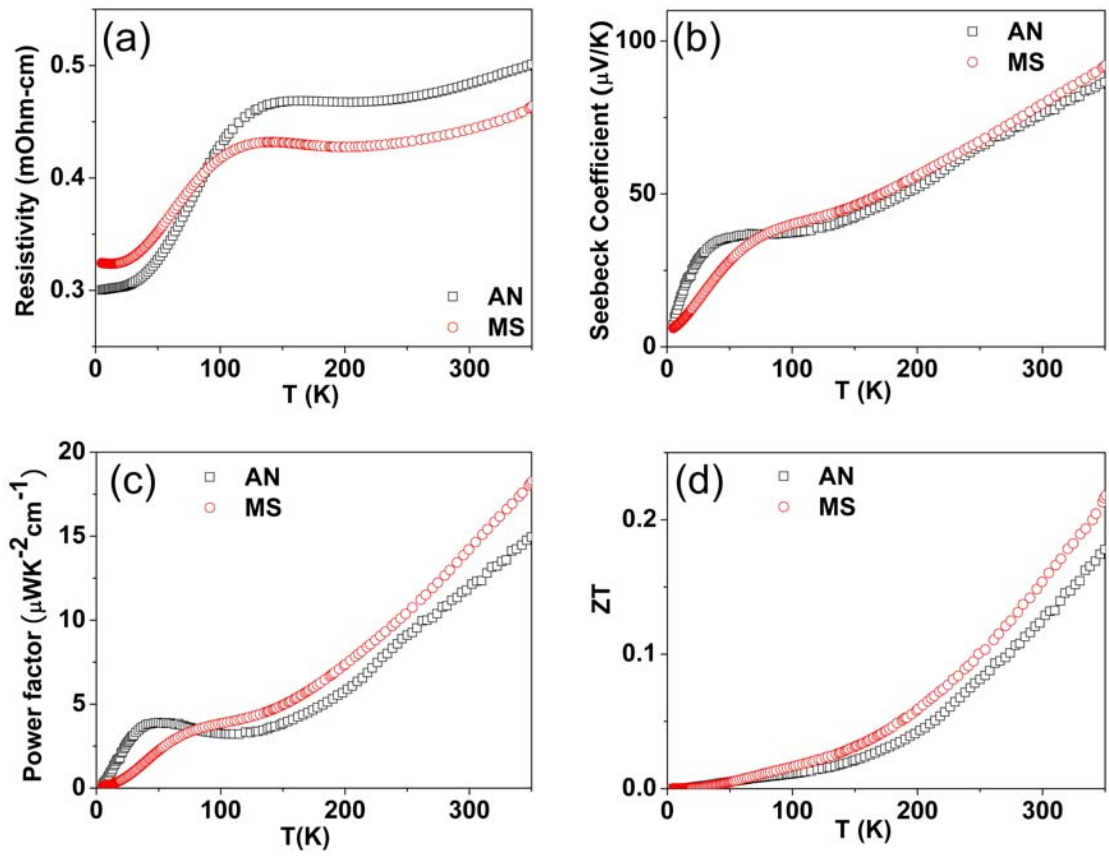

Figure 3

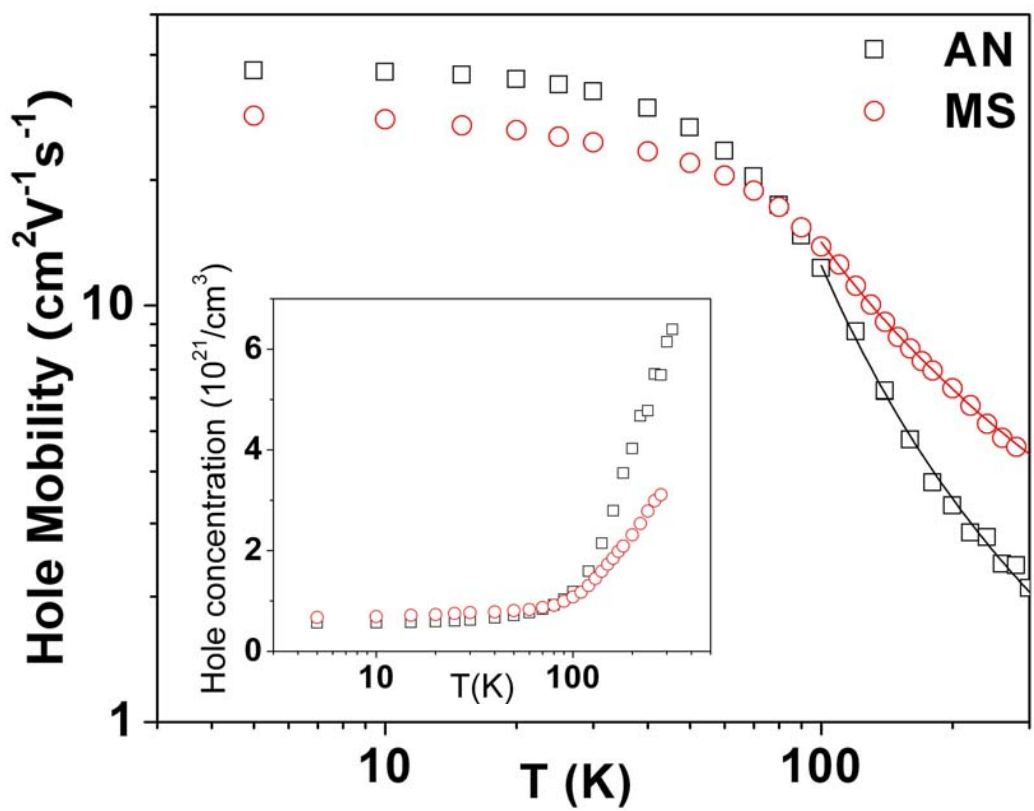

Figure 4